\documentclass[%
 superscriptaddress,
preprint,
 amsmath,amssymb,
aps,
]{revtex4-2}

\usepackage[utf8]{inputenc}
\usepackage{graphicx}
\usepackage{dcolumn}
\usepackage{bm}
\usepackage{xcolor}
\usepackage{hyperref}

\begin{document}

\title{Phase Separation in Wetting Ridges of Sliding Drops on Soft and Swollen Surfaces}

\author{Lukas Hauer}
\altaffiliation{These authors contributed equally}
\affiliation{Max Planck Institute for Polymer Research, Ackermannweg 10, 55128 Mainz, Germany}
\affiliation{ Department of Chemical and Materials Engineering, University of Kentucky, Lexington, KY, USA}

\author{Zhuoyun Cai}
\altaffiliation{These authors contributed equally}
\affiliation{ Department of Chemical and Materials Engineering, University of Kentucky, Lexington, KY, USA}%

\author{Doris Vollmer}
\email{vollmerd@mpip-mainz.mpg.de}
\affiliation{Max Planck Institute for Polymer Research, Ackermannweg 10, 55128 Mainz, Germany}

\author{Jonathan T. Pham}
\email{Jonathan.Pham@uky.edu}
\affiliation{ Department of Chemical and Materials Engineering, University of Kentucky, Lexington, KY, USA}%

\date{\today}

\begin{abstract}
Drops in contact with swollen, elastomeric substrates can induce a capillary-mediated phase separation in wetting ridges. Using laser scanning confocal microscopy, we visualize phase separation of oligomeric silicone oil from a crosslinked silicone network during steady-state sliding of water drops. We find an inverse relationship between the oil tip height and the drop sliding speed, which is rationalized by competing transport timescales of oil molecules: separation rate and drop-advection speed. Separation rates in highly swollen networks are as fast as diffusion in pure melts.

\end{abstract}

\maketitle

\section{Introduction}

Classically, wetting is characterized by the Young-Dupré contact angle between a drop and a substrate at the three-phase-contact line \cite{young_iii_1805}. However, when the underlying substrate is a liquid or a soft solid, this angle does not suffice because of an out-of-plane ridge formation \cite{yuk_contact_1986, schellenberger_direct_2015, semprebon_apparent_2016}. A ‘wetting ridge’ emerges due to the normal component of the drop surface tension, which pulls on the substrate and deforms it upwards \cite{carre_viscoelastic_1996, saiz_ridging_1998, clanet_onset_2002, leonforte_statics_2011, park_selfspreading_2017, gerber_wetting_2019, masurel_elastocapillary_2019, andreotti_statics_2020, dai_droplets_2022, chaudhuri_temperaturedependent_2022}. On pure liquid substrates, the ridge geometry is solely governed by capillarity, while on soft solid substrates, elastic contributions add to the ridge geometry \cite{mora_capillarity_2010,jerison_deformation_2011, bico_elastocapillarity_2018}. The Neumann angles consider force balances not only in the horizontal but also in the vertical direction \cite{neumann_equationofstate_1974}, and can help recover a better description of the wetting situation \cite{marchand_contact_2012, cao_polymeric_2015, gorcum_spreading_2020}. Although wetting of pure liquid or soft solid substrates are typically treated as two distinct cases, many substrates have features of both. For example, crosslinked polymeric substrates are often swollen with unbound, free mobile molecules (e.g. liquid oligomers in elastomers or water in hydrogels \cite{ tanaka_gels_1981, lee_solvent_2003a}); this leads to a complex combination of liquid and solid behaviors. Recently, it has been shown that ridges on swollen, lightly crosslinked elastomers do not necessarily comprise a homogeneous phase. Unbound molecules can phase-separate at the tip of the ridge, forming a region of pure liquid \cite{jensen_wetting_2015,pham_elasticity_2017,cai_fluid_2021}. However, these have been mostly considered in static drops. 

In dynamic wetting conditions, the ridge is highly relevant: Friction that builds up during drop sliding dissipates mostly in the ridge \cite{shanahan_viscoelastic_1995,long_static_1996,karpitschka_droplets_2015,zhao_geometrical_2018}. Hence, the shape and material makeup of the wetting ridge are central components to determine drop movement. For soft, swollen elastomers, both the network and unbound molecules play a role in the wetting ridge. Hence, the presence of free molecules is likely to alter the drop dynamics \cite{hourlier-fargette_role_2017, zhao_growth_2018, xu_viscoelastic_2020}. Yet, it is still not understood how drop sliding speed affects and couples to phase separation, how the sliding-induced separation is related to the swelling ratio of the underlying network, and what time scales govern the separation mechanism.

In this Letter, we investigate wetting ridges on soft and swollen polydimethylsiloxane (PDMS) substrates, formed during steady-state sliding of aqueous drops. Crosslinked PDMS networks are swollen with inert, low molecular weight silicone oil (liquid oligomers), both dyed with individual markers. We monitor the wetting ridge of sliding drops with laser scanning confocal microscopy (LSCM). Cross-sectional views of the moving wetting ridge yield their shape and the spatial distribution of network and oil phases in the ridge. For highly swollen networks, the extent of phase separation depends on the sliding speed of the drop, and is suppressed at fast sliding speeds ($>100~\mathrm{\mu m/s}$). On moderately swollen substrates, however, no phase separation is observed even for slower moving drops. From images of the speed-dependent wetting ridge, we extract the time scale for oil ridge-formation. Tuning the swelling ratio changes the formation timescales across three orders of magnitude.

\section{Material system} Soft PDMS networks ($3-5~\mathrm{kPa}$) with different amounts of swelling agent (\textit{i.e.}, swelling ratio $Q$) were manufactured (details in Supporting Material \footnote{See Supplemental Material [URL will be inserted by publisher] for details on swollen PDMS substrate preparation.} and \cite{glover_extracting_2020}). For this, PDMS networks were initially cleaned of uncrosslinked material followed by reswelling of silicone oil with well-defined low molecular weight (5 cSt/$770~\mathrm{g/mol}$, Gelest). Network and oil are individually labeled with fluorescence markers. The substrate thickness is $\approx 100~\mathrm{\mu m}$, which is large enough to not interfere with the wetting ridge \cite{khattak_direct_2022}. Swollen substrates are placed onto glass slides for drop sliding experiments.

\begin{figure}[t]
\includegraphics[width=0.49\textwidth, angle=0]{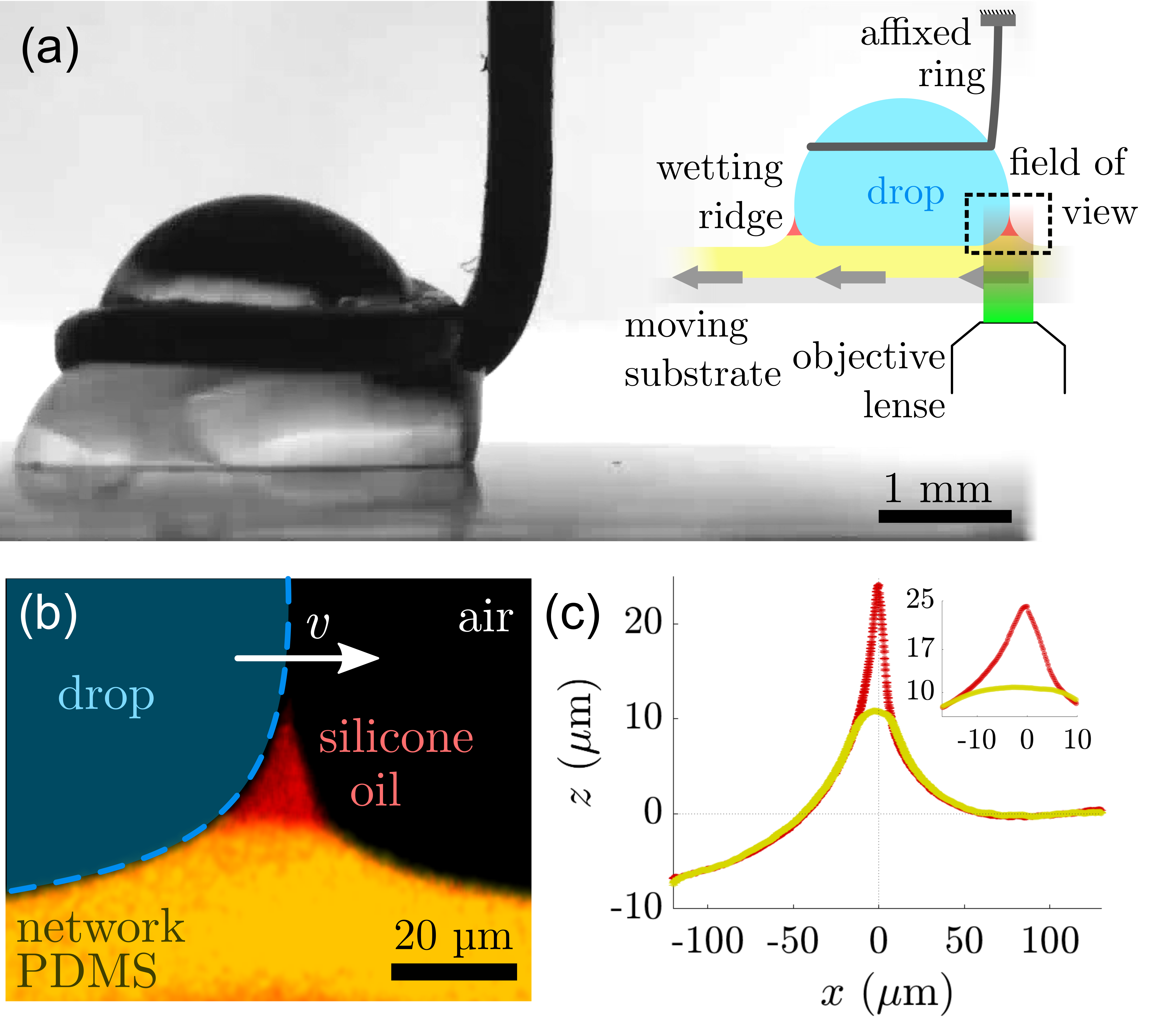}
\caption{\label{fig:fig_1}Drop sliding set-up and wetting ridge visualization. (a) Macroscopic side view of affixed drop, sliding at $5~\mathrm{\mu m/s}$ on a swollen PDMS network ($Q=14.5$). Substrate moves left while drop is stationary. Inset, set-up schematic. (b) Temporal averaged ($n=158$) LSCM image of a phase-separated wetting ridge. PDMS network and silicone oil are dyed separately with fluorescence markers with different emission spectra. Red shows silicone oil and orange shows swollen PDMS network. (c) Extracted interfaces of silicone oil (red points) and PDMS network (yellow points). Inset, blown up section of the phase-separated zone in the wetting ridge. Standard errors are smaller than symbol size.}
\end{figure}

\section{Phase-separated wetting ridge of dynamic drops} The swollen PDMS networks are mounted on a motorized linear stage. A $8~\mathrm{\mu l}$ water drop is then placed on the substrate. Upon deposition, there is an initial drop dwell time, in which an annular wetting ridge forms at the three-phase contact line. An affixed metal ring (diameter $\approx 2.5~\mathrm{mm}$) hovers $1~\mathrm{mm}$ above the substrate and holds the drop in position while the stage moves at constant speeds ($5 - 800~\mathrm{\mu m/s}$), Fig. \ref{fig:fig_1}(a). This generates a relative sliding motion between the drop and substrate, while holding the drop fixed within the laboratory frame for imaging. The setup is mounted on a laser scanning confocal microscope (Leica TCS SP8) that enables microscopic visualization of the wetting ridge. The drop is positioned such that the advancing contact zone lies in the field of view, which spans $250\times 62~\mathrm{\mu m^2}$. While the drop is brought up to its set-point speed, the system shows transient startup dynamics; however, this regime of motion lies outside the scope of our current study on steady-state dynamics. In steady-state motion, the wetting ridge assumes a near-constant shape at a stationary position within the field of view, even during long sliding times ($>200~\mathrm{s}$). We note that the dynamic wetting ridge shape does not depend on the dwell time of the drop prior to sliding. However on rare occasions, the ridge deviates from its stationary position due to contact line pinning, likely stemming from surface impurities or contamination \cite{forsberg_contact_2010,kajiya_advancing_2013,lhermerout_controlled_2018}. Images are taken $1.6\times$ per second, which enables resolving these motions. This fast recording, however, brings higher signal noise with it - a drawback that we overcome by aligning each image by the tip of the ridge, followed by averaging, Supporting Information \footnote{See Supplemental Material [URL will be inserted by publisher] for details for LSCM image processing and source codes.}. 

In a first set of experiments, we slide drops over a swollen substrate ($Q=14.5$) at a speed of $5~\mathrm{\mu m/s}$. We start recording images $20-30~\mathrm{s}$ after the onset of sliding, when no more variations in the ridge shape occur. The dynamic wetting ridge is recorded for $\approx18.5~\mathrm{s}$. This gives $n=158$ images that yield a crisp reconstruction of the averaged wetting zone, Fig. \ref{fig:fig_1}(b). The wetting ridge clearly shows two phases of (i) swollen, network PDMS (orange) and (ii) pure liquid silicone oil (red). While the network height is only slightly elevated, the silicone oil forms a sharp tip. Extracting the interfaces of each phase reveals more quantifiable detail, Fig. \ref{fig:fig_1}(c). The small relative errors of the temporally accumulated data indicates that indeed, the ridge is in a steady-state. At the three-phase contact line, the surface tension of the water pulls up the wetting ridge. At $x=0$, the wetting ridge has its highest point. At $|x| \gg 0$, interfacial profiles of the two phases (red and yellow) align. At $x \ll 0$, the PDMS network profile is dented due to the Laplace pressure in the drop acting on the substrate. The network and the silicone oil are well separated between $-15~\mathrm{\mu m} \lesssim x \lesssim 8~\mathrm{\mu m}$, Fig. \ref{fig:fig_1}(c) inset. We note that within this range, the ridge profile is asymmetric with more silicone oil towards the drop ($x<0$). Additionally, in the separated region, the network profile bends into the oil phase. The positive curvature of the network profile indicates an over-pressure inside the network PDMS with respect to the oil phase. At $x=0$, phase-separation is strongest with a separation height of more than $10~\mathrm{\mu m}$.
\begin{figure*}[t]
\includegraphics[width=1\textwidth, angle=0]{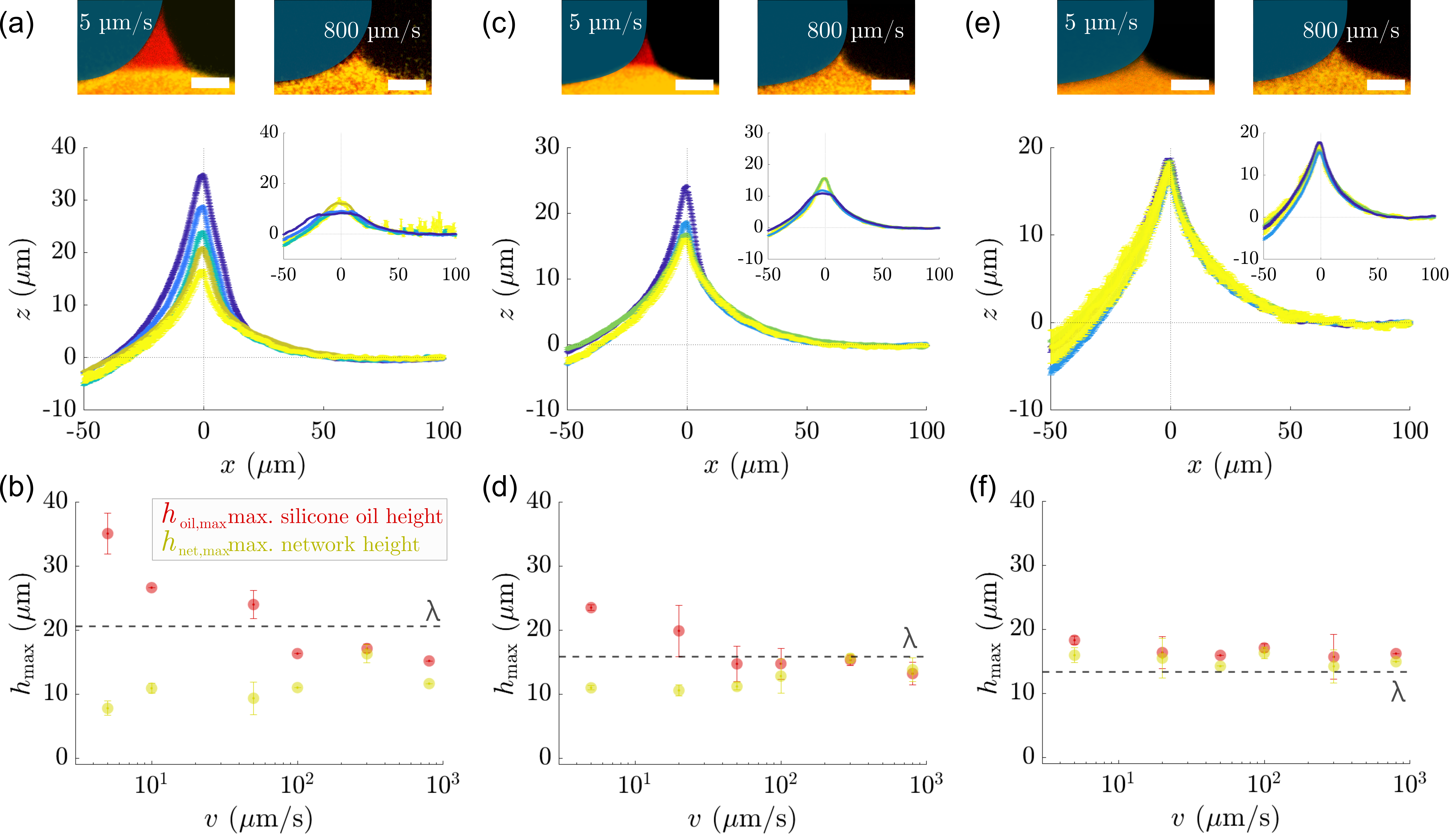}
\caption{\label{fig:fig_2}Dynamic wetting ridge shape for different sliding speeds and swelling ratios (a), (b) $Q=16$, (c), (d) $Q=14.5$, (e), (f) $Q=10$. (a), (c), (e) Dynamic ridge profiles of liquid silicone oil (and network PDMS as insets). (a), (c) For silicone oil, the ridge height gradually decreases for increasing drop speed, $v=5,10,50,100,300,800~\mathrm{\mu m/s}$. Top row images are representative LSCM images for $v=5~\mathrm{\mu m/s}$ and $v=800~\mathrm{\mu m/s}$ (scale bar $20~\mathrm{\mu m}$). (b), (d), (f) Maximum height of dynamic ridges of silicone oil (red) and network PDMS (yellow). Data shows average of min. $n=3$ repetitions together with standard deviations. Dashed lines mark dynamic elastocapillary height $\lambda \sim \gamma\sin{\theta_{\mathrm{adv}}}/E$, which are (b) $21~\mathrm{\mu m}$ ($E\approx3~\mathrm{kPa}$) (d) $16~\mathrm{\mu m}$ ($E\approx3.9~\mathrm{kPa}$) (f) $1.3~\mathrm{\mu m}$ ($E\approx4.8~\mathrm{kPa}$).}
\end{figure*}

\section{Sliding speed variation} 
\begin{figure}
\includegraphics[width=0.49\textwidth, angle=0]{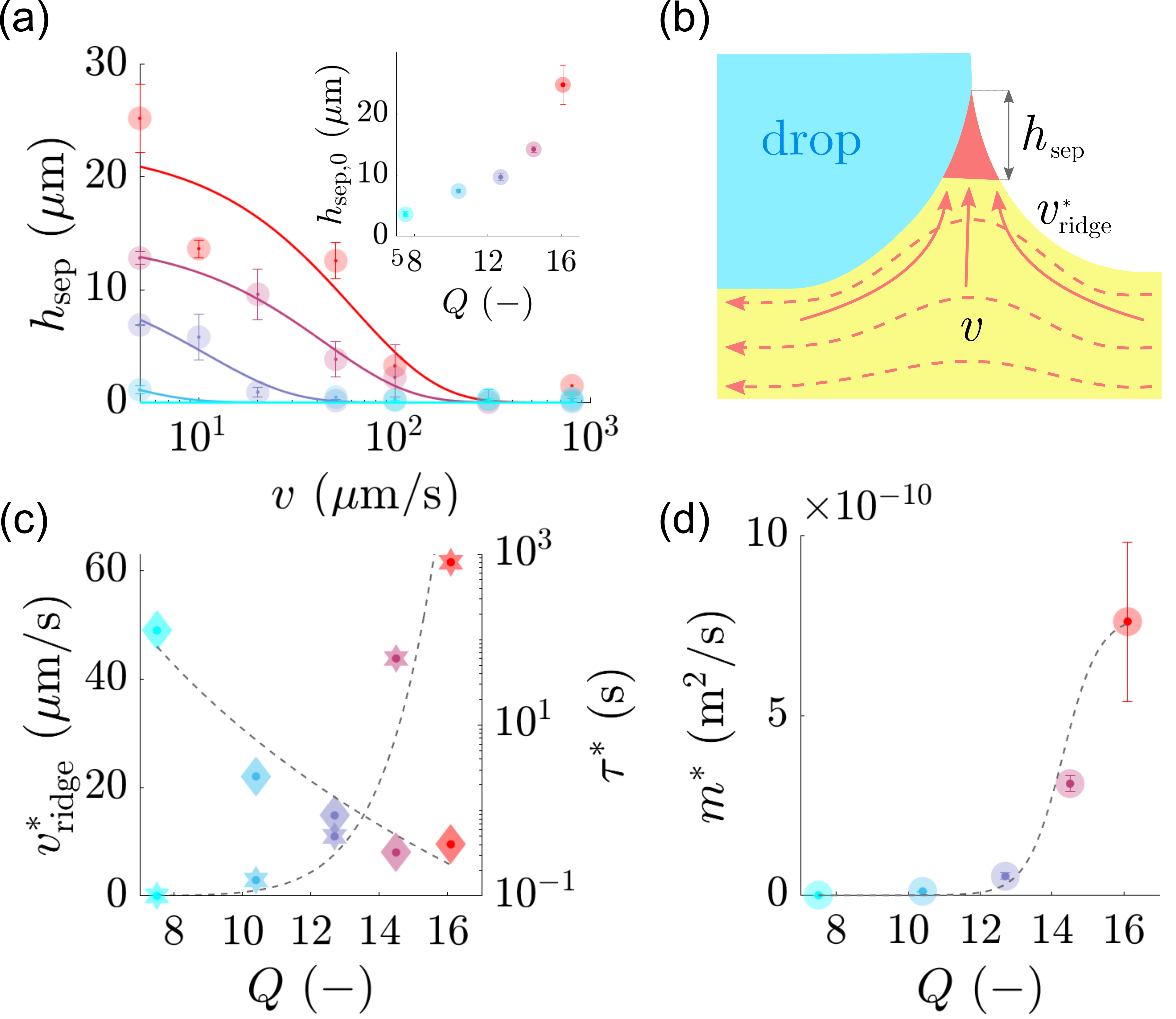}
\caption{\label{fig:fig_3}Separation dynamics of liquid, silicone oil wetting ridge. (a) Separation height $h_{\mathrm{sep}}$ versus $v$, for $Q=16$ (cherry red), $Q=14.5$ (bordeaux), $Q=12.7$ (dark blue), $Q=10$ (blue), and $Q=7.5$ (turquoise). Solid lines are fit to Eq. (\ref{eq:adapted_height}). Inset, extrapolated separation size at zero speed for different $Q$. (b) Competing transport mechanisms of phase separating, silicone oil molecules during drop sliding. Solid lines illustrate the molecular flux $j$ and dashed lines the advective flux. (c) Characteristic ridge growth speed $v^*_\mathrm{ridge}$ (stars) and characteristic growth times $\tau^*_\mathrm{ridge}$ (diamonds) for different $Q$. (d) Measured mobility $m^*$ for different $Q$. Dashed lines in (c) and (d) indicate trends. Error bars for $h_\mathrm{sep,0}$ and $m^*$ correspond to fitted root-square-mean error. In (c), symbol size exceeds error bars.}
\end{figure}

The dynamic ridge profiles of silicone oil are shown in Fig. \ref{fig:fig_2}(a) for $Q=16$ at various sliding speeds, together with the network profiles in the inset. The corresponding averaged LSCM images at the slow and fast speeds are shown in the top row. We varied the speed between $v = 5$ and $800~\mathrm{\mu m/s}$. The highest ridge ($\approx 35~\mathrm{\mu m}$) forms at low sliding speed ($5~\mathrm{\mu m/s}$, dark blue). Increasing $v$ gradually decreases the ridge height. The smallest recorded ridge is less than $20~\mathrm{\mu m}$ high at $v=800~\mathrm{\mu m/s}$ (light yellow). This height-speed dependency is inverted for the network profile, \textit{i.e.}, the network rises with increasing speed. However, the sensitivity of the network height to $v$ is not as pronounced as for the silicone oil tip height. This becomes more evident when plotting the maximum heights of the PDMS network $h_{\mathrm{net,max}}$ and the silicone oil $h_{\mathrm{oil,max}}$ against $v$, Fig. \ref{fig:fig_2}(b). $h_{\mathrm{oil,max}}$ (red circles) decreases much steeper than $h_{\mathrm{net,max}}$ (yellow circles) increases with faster speeds. For $v>100~\mathrm{\mu m/s}$, the height difference between $h_{\mathrm{oil,max}}$ and $h_{\mathrm{net,max}}$ is only $1-3~\mathrm{\mu m}$. Additionally, the ridge becomes narrower with faster $v$. For substrates swollen to a lesser extent ($Q=14.5$), we observe similar speed dependencies, Fig. \ref{fig:fig_2}(c)-(d). Quantitatively, these substrates have smaller ridge heights ($\approx 25~\mathrm{\mu m}$ for $v=5~\mathrm{\mu m/s}$) and narrower widths compared to the saturated ($Q=16$) substrate. The less swollen substrate also has a less sensitive height-speed dependency. For $v>100~\mathrm{\mu m/s}$, the silicone oil does not clearly separate. When substrates are swollen even less to $Q=10$, the ridges display no speed dependency and no visual phase separation; within experimental accuracy, all dynamic ridges collapse to the same shape, \textit{i.e.}, $h_{\mathrm{oil,max}} \approx h_{\mathrm{net,max}}$, Fig. \ref{fig:fig_3}(e). The maximum ridge height reaches about $15~\mathrm{\mu m}$ for the entire speed spectrum. Thus, the drop sliding speed is a critical factor that governs phase separation in dynamic wetting.\\
\indent To better understand the dynamic conditions for when phases separate, we relate the network height to the force acting normal to the ridge. The network PDMS is (visco)elastic and adapts its shape to imposed stresses. Therefore, $h_{\mathrm{net, max}}$ is a direct proxy for the force $f$, acting on the material per 
\begin{equation}
    \label{eq:elast_length}
   h_{\mathrm{net,max}} \sim f/E.
\end{equation}
Here, $E$ is the Young's modulus of the network measured previously in \cite{cai_fluid_2021}. On PDMS networks, without phase separation, the force is imposed at a singular point by the capillary action of the drop, namely
\begin{equation}
    \label{eq:capillar_force}
    f \sim \gamma\sin{\theta_{\mathrm{adv}}}
\end{equation}
where $\gamma$ is the drop surface tension and $\theta_{\mathrm{adv}}$ the advancing contact angle. In the case of no phase separation, $h_{\mathrm{net,max}}$ should coincide with the advancing elastocapillary length 
\begin{equation}
    \label{eq:elastocapilary_length}
    \lambda \sim \gamma\sin{\theta_{\mathrm{adv}}}/E.    
\end{equation}
 Swollen PDMS shows advancing contact angles of $\theta_{\mathrm{adv}}\approx 105^\circ$ \cite{cai_how_2022}. Liquid silicone oil tends to cloak aqueous drops and consequently lowers $\gamma$ from $72~\mathrm{mN/m}$ to $64~\mathrm{mN/m}$ \cite{hourlier-fargette_extraction_2018,naga_how_2021}. When no pronounced phase separation occurs, $\lambda$ and $h_{\mathrm{net,max}}$ coincide indeed reasonably well, Fig. \ref{fig:fig_2}(b),(d) and (f) dashed line. Discrepancies may arise from small measurement errors of $E$. For substrates of $Q>10$, the transition to suppressed phase separation happens at speeds $v>100~\mathrm{\mu m/s}$. On $Q<10$, both heights align for all $v$ ($>5~\mathrm{\mu m/s}$), indicating that phase separation is mostly suppressed. For $Q=16$ and $14.5$, $h_{\mathrm{net,max}}$ falls at slower speeds. The height-speed trend inverses for $h_{\mathrm{oil,max}}$. This indicates a coupling between $h_{\mathrm{oil,max}}$ and $h_{\mathrm{net,max}}$: The normal force acting on the network relaxes from the initial capillary induced (singularity) force, Eq. (\ref{eq:capillar_force}), over the region of phase separation. Alternatively, $h_{\mathrm{net,max}}$ may also change due to altered material composition (\textit{i.e.}, $E$) in the network ridge when phases separate. However, we expect that the material composition in the network remains largely unaffected by variations in $v$, due to the large reservoir of silicone oil in the bulk. \ 

\section{Dynamic separation height} The separation height of the liquid silicone oil ridge is the difference of the total height and the network height, \textit{i.e.}, $h_{\mathrm{sep}}=h_{\mathrm{oil,max}} - h_{\mathrm{net,max}}$. To understand how it relates to sliding speed, we plot $h_{\mathrm{sep}}$ against $v$ for substrates swollen to $Q=16,14.5,12.7,10,$ and $7.5$, Fig. \ref{fig:fig_3}(a). For substrates swollen to $Q>10$, we consistently observe phase-separated ridges for $5~\mathrm{\mu m/s}<v<100~\mathrm{\mu m/s}$. In this range, $h_{\mathrm{sep}}$ falls monotonously with increasing $v$.\

 To understand the relation between $h_{\mathrm{sep}}$ and $v$, we consider the formation process of the phase-separated ridge while the drop slides over the substrate. Thermodynamically, the oil ridge formation is driven by a chemical potential gradient $\nabla \mu$ between the network and the oil phases. The resulting molecular flux is $j = -m\nabla \mu,$ where $m$ is the mobility of the oil molecules. Given the poroelastic nature of the flux \cite{biot_general_1941, hu_using_2010, berman_singular_2019}, the mobility can be thought of as a Darcy type, $m=k/\eta \Omega^2$, where $k$ is the network permeability, $\eta = 4.6 ~\mathrm{mPa~s}$ the viscosity, and $\Omega=840~\mathrm{mL/mol}$ the molar volume of the low molecular weight silicone oil \cite{polydimethylsiloxane_}. Corresponding to $j$, we can define a characteristic molecular speed that is related to the separated ridge formation, $v^*_\mathrm{ridge}$. During sliding, the velocity of the oil molecules in the network is not only defined by $v^*_\mathrm{ridge}$, but superposed by the advective speed of the sliding drop $v$, Fig \ref{fig:fig_3}(b). These two speeds are in competition, effectively governing the dynamic, oil separation height. When $v\gg v^*_\mathrm{ridge}$, oil molecules do not have time to travel to the ridge, but pace below the drop. As a consequence, the molecules stay in the network and the separation height $h_{\mathrm{sep}}$ remains small. The phase-separated ridge can only fully emerge when $v\ll v^*_\mathrm{ridge}$. In between these two limits, the separation height will depend on the sliding speed. We formulate this height-speed relation as a first order process, namely
 \begin{equation}
 \label{eq:adapted_height}
     h_{\mathrm{sep}} = h_{\mathrm{sep,}0} - \Delta h_{\mathrm{sep,dyn}} \left[1-e^{-v/v^*_{\mathrm{ridge}}}\right],
 \end{equation}
 
\noindent where $h_{\mathrm{sep,}0}$ is the separation height at zero speed and $\Delta h_{\mathrm{sep,dyn}}$ is the maximum magnitude of ridge separation. When $v$ rises, $h_{\mathrm{sep}}$ adapts to the dynamic state of the drop and gradually decreases. $\Delta h_{\mathrm{sep,dyn}}$ depends weakly on the range of investigated speeds; in the limit of $v\gg v^*_\mathrm{ridge}$, $\Delta h_{\mathrm{sep,dyn}}$ should converge to $h_{\mathrm{sep,}0}$. This framework enables us to extract the zero-speed height $h_{\mathrm{sep,}0}$, and the characteristic formation speed $v^*_{\mathrm{ridge}}$, Fig. \ref{fig:fig_3}(a)-(c). Extracted values (inset in Fig 3a) for $h_{\mathrm{sep,}0}$ are slightly higher than the measured values of $h_{\mathrm{sep}}$ at $v=5~\mathrm{\mu m/s}$, and coincide well with static measurements \cite{cai_fluid_2021}. Extracted values for $v^*_{\mathrm{ridge}}$ are diminishingly small for $Q<10$, which aligns with the observation that no pronounce phase separation occurs. For $Q>10$, $v^*_{\mathrm{ridge}}$ becomes significantly larger than zero, \textit{i.e.}, $\mathcal{O}\left( v^*_{\mathrm{ridge}} \right)= 10~\mathrm{\mu m/s}$. $v^*_{\mathrm{ridge}}$ is linked to the migration speed of oil molecules inside the network. Hence, a characteristic time scale of the migrating molecules is related as 
  \begin{equation}
     \tau ^* \sim h_{\mathrm{sep,}0}/v^*_{\mathrm{ridge}}.
 \end{equation}
 
 From the data in Fig. \ref{fig:fig_3}(c), we compute the time scales of migrating oil molecules. Notably, for $Q>10$, $\tau^*$ remains mostly constant ($0.3-0.4~\mathrm{s}$). With the characteristic time ($\tau^*$) and length ($h_{\mathrm{sep,}0}$) scales, we can now estimate effective mobilities as
 \begin{equation}
     m^*\sim h_{\mathrm{sep,}0}^2 / 2 \tau^*,
 \end{equation}
plotted in Fig. \ref{fig:fig_3}(d). For $Q<10$, $m^*\approx 0$ since we do not observe any phase-separated region and therefore no molecular flux builds up. For $Q>10$, the mobility increases with swelling ratio and $\mathcal{O}\left( m^* \right)=10^{-10}~\mathrm{m^2/s}$, which compares well with mobilities of self diffusing (low $\Omega$) PDMS molecules in melts \cite{mccall_self_1965}. For $Q>10$, the molecules become gradually more mobile. This can be explained by considering the network structure: When more oil is swollen into the network, the pores of the network (\textit{i.e.}, mesh size) are expanded. When molecules travel through the expanded pores, the imposed friction from the immobilised (crosslinked) network is reduced. This is reflected in higher permeability values $k$, and hence, a higher Darcy mobility. At a given prevailing chemical potential gradient, combined with an inherent excess in available oil molecules, high swelling ratios lead to faster molecular fluxes, and ultimately, to faster and larger ridge formations. Eventually, the imposed friction in strongly expanded networks stems dominantly from internal molecular friction, since the interactions with the immobile network becomes negligible. Hence at high swelling ratio, transport crosses over from a Darcy type to pure melt diffusion.

In summary, the competition of drop advection and molecular flux governs the degree of phase separation in wetting ridges of moving drops. Understanding phase separation should offer guidelines on controlling drop dynamics. It is notable that phase separation is locally triggered by the (weak) singularity of the three-phase contact line. Since phase separation is fully suppressed at certain drop speed and swelling combinations, it raises the question of whether this can be considered as a first-order phase transition. 

\section{Acknowledgements}

We thank Rodrique Badr, Abhinav Naga, and William S.Y. Wong for discussions. This research is supported by the US National Science Foundation through award number 2043732 (Z.C., J.T.P.), German Research Foundation (DFG) with the Priority Programme 2171 (L.H., D.V.).

\bibliographystyle{apsrev4-1}
\bibliography{aipsamp}

\end{document}